\documentclass[12pt]{article}
\begin{document}

\def \d {{\rm d}}

\title{Nonexpanding impulsive gravitational waves with an arbitrary
cosmological constant}

\author{J. Podolsk\'y\thanks{E--mail: {\tt Podolsky@mbox.troja.mff.cuni.cz}}
\\
\\ Department of Theoretical Physics, Charles University,\\
V Hole\v{s}ovi\v{c}k\'ach 2, 18000 Prague 8, Czech Republic.\\
\\
and J. B. Griffiths\thanks{E--mail: {\tt J.B.Griffiths@Lboro.ac.uk}} \\ \\
Department of Mathematical Sciences, Loughborough University \\
Loughborough, Leics. LE11 3TU, U.K. \\ } 

\date{\today}

\maketitle
\begin{abstract}
Exact solutions for nonexpanding impulsive waves in a background with nonzero
cosmological constant are constructed using a ``cut and paste'' method. These
solutions are presented using a unified approach which covers the cases of
de~Sitter, anti-de~Sitter and Minkowski backgrounds. The metrics are
presented in continuous and distributional forms, both of which are conformal
to the corresponding metrics for impulsive {\sl pp}-waves, and for which the
limit as $\Lambda\to0$ can be made explicitly. 
\end{abstract}


Penrose \cite{Pen72} has presented a geometrical method for the construction
of plane (non\-expanding) and spherical (expanding) impulsive gravitational
waves in a Minkowski background by cutting the space-time along a null
hypersurface and then re-attaching the two pieces with a suitable warp. The
first case leads to impulsive {\sl pp}-waves. The second case describing
expanding impulsive waves \cite{GlePul89}--\cite{Hogan93} has been
extended to backgrounds with a nonzero cosmological constant $\Lambda$
\cite{Hogan92}. However, the method has not previously been explicitly used
for the construction of nonexpanding waves in backgrounds with $\Lambda\ne0$,
although such solutions are already known \cite{HotTan93}--\cite{Podol98}.
The purpose of the present letter is to derive these solutions using the
Penrose method. Remarkably, this approach leads to a convenient unified
representation of the whole family of solutions in which it is possible to
set $\Lambda=0$ explicitly.

Let us first recall the line element for a space-time of constant curvature
in the manifestly conformally flat form 
 \begin{equation}
\d s_0^2= {2\d u\,\d v -2\d\zeta\d\bar\zeta
\over [1-{1\over6}\Lambda(uv-\zeta\bar\zeta)]^2},
 \label{deS}
 \end{equation}
 where $\Lambda$ is the cosmological constant. This is de~Sitter space when
$\Lambda>0$, anti-de~Sitter space when $\Lambda<0$ and Minkowski space when
$\Lambda=0$. As is well known, the (anti-)de~Sitter space-time can be
represented as the 4-dimensional hyperboloid 
 $$ {Z_0}^2-{Z_1}^2-{Z_2}^2-{Z_3}^2-\epsilon{Z_4}^2=-\epsilon a^2, \qquad
a^2={3\over\epsilon\Lambda}, $$ 
 in the flat 5-dimensional space \ 
 $\d s_0^2=\d{Z_0}^2-\d{Z_1}^2-\d{Z_2}^2-\d{Z_3}^2-\epsilon\d{Z_4}^2$, \
where $\epsilon=1$ for $\Lambda>0$ and $\epsilon=-1$ for $\Lambda<0$. The
coordinates of the metric (\ref{deS}) form a suitable parameterization of the
hyperboloid in which 
 \begin{eqnarray}
 Z_0 &=& {\textstyle{1\over\sqrt2}(u+v)
\left[1-{1\over6}\Lambda(uv-\zeta\bar\zeta)\right]^{-1}}, \nonumber\\ 
 Z_1 &=& {\textstyle{1\over\sqrt2}(u-v)
\left[1-{1\over6}\Lambda(uv-\zeta\bar\zeta)\right]^{-1}}, \nonumber\\ 
 Z_2 &=& {\textstyle{1\over\sqrt2}(\zeta+\bar\zeta)
\left[1-{1\over6}\Lambda(uv-\zeta\bar\zeta)\right]^{-1}}, \label{Zcoords}\\ 
 Z_3 &=& {\textstyle -i{1\over\sqrt2}(\zeta-\bar\zeta)
\left[1-{1\over6}\Lambda(uv-\zeta\bar\zeta)\right]^{-1}}, \nonumber\\ 
 Z_4 &=& \sqrt{3\over|\Lambda|}\>
\left[{1+{1\over6}\Lambda(uv-\zeta\bar\zeta)
\over1-{1\over6}\Lambda(uv-\zeta\bar\zeta)}\right]. \nonumber
 \end{eqnarray}
 Inversely, this is given by $u={1\over\sqrt2}(Y_0+Y_1)$,
$v={1\over\sqrt2}(Y_0-Y_1)$, $\zeta={1\over\sqrt2}(Y_2+iY_3)$ where
$Y_\alpha=2a\,Z_\alpha/(Z_4+a)$ with $\alpha=0,1,2,3$. It may be observed
that these coordinates cover the complete hyperboloid, although there is a
coordinate singularity along the section $Z_4=-a$.

Let us now consider the transformation $u=U$, $v=V+H+UH_{Z}H_{\bar Z}$,
$\zeta=Z+UH_{\bar Z}$ applied to the line element (\ref{deS}), where
$H=H(Z,\bar Z)$ is an arbitrary real function. This results in the metric  
 \begin{equation}
\d s_0^2= {2\d U\,\d V 
-2|\d Z+U(H_{Z\bar Z}\d Z+H_{\bar Z\bar Z}\d\bar Z)|^2
\over [1-{1\over6}\Lambda(UV-Z\bar Z+UG)]^2},
 \label{deS2}
 \end{equation}
 where $G=H-ZH_Z-\bar ZH_{\bar Z}$.

Following Penrose's ``cut and paste'' method \cite{Pen72}, we may now take
the line element (\ref{deS}) with $u=U$, $v=V$ and $\zeta=Z$ for $U<0$ and
combine this with (\ref{deS2}) for $U>0$. The resulting line element 
 \begin{equation}
\d s^2= {2\d U\,\d V 
-2|\d Z+U\Theta(U)(H_{Z\bar Z}\d Z+H_{\bar Z\bar Z}\d\bar Z)|^2
\over [1-{1\over6}\Lambda(UV-Z\bar Z+U\Theta(U)G)]^2},
 \label{cont}
 \end{equation}
 where $\Theta(U)$ is the Heaviside step function, is continuous across the
null hypersurface $U=0=u$. However, the discontinuity in the derivatives of
the metric yields impulsive components in the curvature tensor proportional
to the Dirac $\delta$-function. These are interpreted as impulsive waves in
de~Sitter, anti-de~Sitter or Minkowski backgrounds. For $\Lambda=0$, this
reduces to the well known Rosen form for impulsive {\sl pp}-waves
\cite{PodVes98}, \cite{KunSte99}.

We may observe from (\ref{deS}) that the geometry of the wavefront $u=0$ is
described by the 2-metric 
 \begin{equation}
\d\sigma^2= -{2\,\d\zeta\,\d\bar\zeta
\over [1+{1\over6}\,\Lambda\,\zeta\,\bar\zeta]^2}, \label{2surface}
 \end{equation}
 which is a 2-dimensional space of constant gaussian curvature $K=\Lambda/3$.
When $\Lambda=0$ the impulsive wave surface is a plane, for $\Lambda>0$ it is
a sphere, while for $\Lambda<0$ it is a hyperboloid. For $\Lambda\ne0$, the
geometry of these surfaces has been described in detail in~\cite{PodGri97}.

Now, the explicit form of the complete transformation formed by
combining those above is given by 
 \begin{eqnarray}
u&=&U, \nonumber\\ 
v&=&V+H\,\Theta(U)+U\,\Theta(U)\,H_{Z}H_{\bar Z},  \label{trans}\\
\zeta&=&Z+U\,\Theta(U)\,H_{\bar Z}. \nonumber
 \end{eqnarray}
 This is discontinuous at $u=0$ in such a way that 
 $$ (u=0,v,\zeta,\bar\zeta)_{M^-}
=(u=0,v-H(\zeta,\bar\zeta),\zeta,\bar\zeta)_{M^+}\>, $$ 
 which is exactly the Penrose junction condition for reattaching the two
halves of the space-time $M^-(u<0)$ and $M^+(u>0)$ with a ``warp''.

Significantly, the transformation (\ref{trans}), taking into account the
terms which arise from the derivatives of $\Theta(u)$, relates the continuous
form of the impulsive wave metric (\ref{cont}), not to the initial metric
(\ref{deS}), but to the following metric which also includes an impulsive
component explicitly located on the wavefront $u=0$
 \begin{equation}
\d s^2= {2\d u\,\d v -2\d\zeta\,\d\bar\zeta
-2H(\zeta,\bar\zeta)\,\delta(u)\,\d u^2 
\over [1-{1\over6}\Lambda(uv-\zeta\bar\zeta)]^2}. 
 \label{confpp}
 \end{equation} 
 This represents a nonexpanding impulsive wave in any background space-time
of constant curvature.

The above result is well known for impulsive waves in a Minkowski background,
where it is exactly the standard Brinkmann form for a general impulsive {\sl
pp}-wave. However, the above form has not previously been given explicitly
for the case when $\Lambda\ne0$. For $\Lambda<0$, an equivalent form (in
$d$-dimensions) has been used in \cite{HorItz99}.

It may be observed that the metric (\ref{confpp}) is conformal to the general
impulsive {\sl pp}-wave. In this context, we recall that Siklos
\cite{Siklos85} has proved that Einstein spaces conformal to {\sl pp}-waves
only occur when $\Lambda<0$. However, it would appear that the {\it
impulsive} case is a counter-example to this result.

In fact, the metric (\ref{confpp}) is a suitable parameterisation of a class
of nonexpanding impulsive wave solutions described in a 5-dimensional
formalism in \cite{PodGri98}: 
 $$ \d s^2= \d{Z_0}^2 -\d{Z_1}^2 -\d{Z_2}^2 -\d{Z_3}^2 
-\epsilon\d{Z_4}^2 -\tilde H(Z_2,Z_3,Z_4)\delta(Z_0+Z_1)(\d Z_0+\d Z_1)^2. $$ 
 Indeed, using (\ref{Zcoords}), we obtain exactly (\ref{confpp}) where 
 \begin{equation}
H(\zeta,\bar\zeta) ={\textstyle{1\over\sqrt2}\,
\left(1+{1\over6}\,\Lambda\,\zeta\,\bar\zeta\right)\, \tilde
H(\zeta,\bar\zeta)}.
 \label{Htrans}
 \end{equation} 
 In this relation, the parameterisation (\ref{Zcoords}), restricted to the
impulsive wave surface $u=0$, is used to express the arguments of $\tilde H$
in terms of $\zeta$ and $\bar\zeta$ only.

The above solutions can describe impulsive gravitational waves or impulses of
null matter. Using the tetrad frame $\ell^\mu=\Omega\,\delta_2^\mu$,
$m^\mu=\Omega{1\over\sqrt2}(\delta_3^\mu+i\delta_4^\mu)$,
$n^\mu=\Omega\,(\delta_1^\mu+H\delta(u)\delta_2^\mu)$ where
$\Omega=1-{1\over6}\Lambda(uv-\zeta\bar\zeta)$, the nonzero components of the
Weyl and Ricci tensors are 
 \begin{eqnarray}
 \Psi_4 &=& -{\textstyle \left(1+{1\over6}\Lambda\,\zeta\,\bar\zeta\right)^2
H_{\zeta\zeta} \,\delta(u), } \nonumber\\
 \Phi_{22} &=& -{\textstyle \left(1+{1\over6}\Lambda\,\zeta\,\bar\zeta\right)
\left[ \left(1+{1\over6}\Lambda\,\zeta\,\bar\zeta\right) H_{\zeta\bar\zeta}
+{1\over6}\Lambda \left(H-\zeta H_\zeta-\bar\zeta H_{\bar\zeta}\right)
\right] \delta(u). }
\nonumber 
 \end{eqnarray}
 With (\ref{Htrans}) the vacuum field equations $\Phi_{22}=0$ can then be
expressed as
 \begin{equation}
\left(1+{\textstyle{1\over6}}\,\Lambda\,\zeta\,\bar\zeta\right)^2 \tilde
H_{\zeta\bar\zeta} +{\textstyle{1\over3}}\,\Lambda\,\tilde H=0,
 \label{vacuum}
 \end{equation}
 which is simply $(\Delta+{2\over3}\,\Lambda)\tilde H=0$, where $\Delta$ is
the Laplacian operator on a 2-dimensional impulsive wave
surface~(\ref{2surface}). This generalises the well known vacuum field
equations for impulsive waves in a Minkowski background. A general solution
of this equation (\ref{vacuum}) is 
 $$ \tilde H(\zeta,\bar\zeta) =(f_\zeta+\bar f_{\bar\zeta})
-{\Lambda\over3}\,{\bar\zeta f+\zeta\bar f\over
\left(1+{1\over6}\,\Lambda\,\zeta\,\bar\zeta\right)}, $$ 
 where $f(\zeta)$ is an arbitrary function of $\zeta$. Thus, the general
vacuum solution for the metric (\ref{confpp}) is given by
$\sqrt2\,H(\zeta,\bar\zeta)= (1+{1\over6}\,\Lambda\,\zeta\,\bar\zeta) 
(f_\zeta+\bar f_{\bar\zeta}) -{1\over3}\,\Lambda\,(\bar\zeta f+\zeta\bar f)$.
This describes an impulsive gravitational wave in which
$\Psi_4=-{1\over\sqrt2} (1+{1\over6}\,\Lambda\,\zeta\,\bar\zeta)^3
f_{\zeta\zeta\zeta} \,\delta(u)$. The space-time is conformally flat
everywhere when the function $f$ is at most quadratic in $\zeta$. However,
solutions of (\ref{vacuum}) necessarily contain singularities. These are
located on the wavefronts and may be considered as null sources of the
impulsive gravitational waves.

For example, the Aichelburg--Sexl solution \cite{AicSex71} obtained by
boosting the Schwarzschild metric is given by \
$f_0={1\over2}\zeta(\log\zeta-1)$. \ Similarly, the Hotta--Tanaka
solution \cite{HotTan93} obtained by boosting the
Schwarzschild--(anti-)de~Sitter metric is given by \
$f_0={1\over2}\zeta(\log\zeta+{1\over2}\log{1\over6}|\Lambda|)$. \ For these
 \begin{eqnarray}
 \Lambda=0: \qquad &&\sqrt2\,H_0 ={\textstyle{1\over2}}\log\zeta\bar\zeta,
\nonumber\\
 \Lambda\ne0: \qquad &&\sqrt2\,H_0
={\textstyle{1\over2}(1-{1\over6}\Lambda\zeta\bar\zeta)
\log({1\over6}|\Lambda|\zeta\bar\zeta) +(1+{1\over6}\Lambda\zeta\bar\zeta)}.
\nonumber
 \end{eqnarray}
 In the second expression, the limit as $\Lambda\to0$ differs from the first
only by a constant term that can be transformed away.  These solutions
describe gravitational waves generated by null monopole point particles in
backgrounds of constant curvature. Further explicit solutions representing
impulsive gravitational waves generated by null particles with arbitrary
multipole structure in these backgrounds have been given previously in
\cite{GriPod97} and~\cite{PodGri98}. For dipole and quadrupole sources, the
solutions are given by \ $f_1=(1-{1\over6}\Lambda\zeta^2)\log\zeta
+\log{1\over6}|\Lambda|$ \ and \
$f_2=-\zeta^{-1}+{1\over36}\,\Lambda^2\,\zeta^3$, \ for which
 \begin{eqnarray}
 &&\sqrt2\,H_1=\left({1\over\zeta}+{1\over\bar\zeta}\right)
(1-[{\textstyle{1\over6}\Lambda\zeta\bar\zeta]^2)
-{1\over3}\Lambda(\zeta+\bar\zeta)
\log({1\over6}|\Lambda|\zeta\bar\zeta)}, \nonumber\\
 &&\sqrt2\,H_2=\left({1\over\zeta^2}+{1\over\bar\zeta^2}\right)
(1+{\textstyle{1\over6}}\Lambda\zeta\bar\zeta)^3, \nonumber
 \end{eqnarray}
 respectively. It may be noted that, for these and higher multipole terms,
the limit for the metric components as $\Lambda\to0$ can be performed
explicitly. This further demonstrates the suitability of these coordinates
for discussing the above class of impulsive solutions in a unified way.

\section*{Acknowledgments}

This work was supported by a visiting fellowship from the Royal Society and,
in part, by the grant GACR-202/99/0261 of the Czech Republic.


\begin{thebibliography}{99}

\bibitem{Pen72} R. Penrose, {\sl General Relativity}
ed L O'Rai\-feartaigh (Clarendon, Oxford, 1972) 101

\bibitem{GlePul89} R. Gleiser and J. Pullin, {\sl Class. Quantum Grav.}
{\bf 6} (1989) L141 

\bibitem{NutPen92} Y. Nutku and R Penrose, {\sl Twistor
Newsletter} No. 34, 11 May (1992) 9

\bibitem{Hogan93} P. A. Hogan, {\sl Phys. Rev. Lett.} {\bf 70} (1993) 117

\bibitem{Hogan92} P. A. Hogan, {\sl Phys. Lett.} A {\bf 171} (1992) 21 

\bibitem{HotTan93} M. Hotta and T. Tanaka, {\sl Class. Quantum Grav.} {\bf
10} (1993) 307  

\bibitem{PodGri98} J. Podolsk\'y and J. B. Griffiths, {\sl Class. Quantum
Grav.} {\bf 15} (1998) 453 

\bibitem{Podol98} J. Podolsk\'y, {\sl Class. Quantum Grav.} {\bf 15} (1998)
3229 

\bibitem{PodVes98} J. Podolsk\'y and K. Vesel\'y, {\sl Phys. Lett.} A
{\bf 241} (1998) 145

\bibitem{KunSte99} M. Kunzinger and R. Steinbauer, {\sl Class. Quantum Grav.}
{\bf 16} (1999) 1255

\bibitem{PodGri97} J. Podolsk\'y and J. B. Griffiths, {\sl Phys. Rev.} D
{\bf 56} (1997) 4756 

\bibitem{HorItz99} G. T. Horowitz and N. Itzhaki, {\sl JHEP} {\bf 9902}
(1999) 010

\bibitem{Siklos85} S. T. C. Siklos, {\sl Galaxies, axisymmetric systems
and relativity} ed M A H MacCallum (Cambridge, 1985) 247 

\bibitem{AicSex71} P. C. Aichelburg and R. U. Sexl, {\it Gen. Rel. Grav.}
{\bf 2} (1971) 303 

\bibitem{GriPod97} J. B. Griffiths and J. Podolsk\'y, {\sl Phys. Lett.} A
{\bf 236} (1998) 8 

\end{thebibliography}
\end{document}